\documentclass[preprint,12pt]{revtex4}

\usepackage{amsmath} 
\usepackage{amsfonts}  
\usepackage{amssymb}
\usepackage{graphicx}  
\usepackage{color}
\newcommand{\ii}{\mbox{i}}
\newcommand{\iii}{\mbox{\footnotesize i}}
\newcommand{\bra}[1]{\langle #1 | \,}
\newcommand{\ket}[1]{\, | #1 \rangle}

\begin{document}




\title{Detection of avoided crossings by fidelity}


\author{Patrick Pl\"otz}
\affiliation{Institut f\"ur Theoretische Physik, Philosophenweg 19, Universit\"at Heidelberg, 69120 Heidelberg, Germany}
\author{Michael Lubasch}
\affiliation{Institut f\"ur Theoretische Physik, Philosophenweg 19, Universit\"at Heidelberg, 69120 Heidelberg, Germany}
\affiliation{Max-Planck-Institut f\"ur Quantenoptik, Hans-Kopfermann-Str. 1, 85748 Garching, Germany}
\author{Sandro Wimberger}
\affiliation{Institut f\"ur Theoretische Physik, Philosophenweg 19, Universit\"at Heidelberg, 69120 Heidelberg, Germany}         
\affiliation{Center for Quantum Dynamics, Philosophenweg 12, Universit\"at Heidelberg, 69120 Heidelberg, Germany}

\begin{abstract}
The fidelity, defined as overlap of eigenstates of two slightly different Hamiltonians, is proposed as an efficient detector of avoided crossings in the energy spectrum. This new application of fidelity is motivated for model systems, and its value for analyzing complex quantum spectra is underlined by applying it to a random matrix model and a tilted Bose--Hubbard system.
\end{abstract}



\maketitle

\section{Introduction}
\label{intro}

The progress in cooling and manipulating ultracold atomic gases in recent years has opened new perspectives on interacting many-body models from condensed matter physics~\cite{BlochZwergerReview,ari}. It led to questions and opportunities beyond conventional solid-state physics, e.g., the direct experimental study of quantum phase transitions~\cite{BlochZwergerReview}, the role and engineering of genuine quantum correlations~\cite{BlochZwergerReview,dirk}, and the phenomenon of quantum chaos in systems that consist of indistinguishable particles~\cite{manybodychaos,FHM,KB2,KB1,TMW2}. In this context, it is possible to detect a quantum phase transition by the change of fidelity (modulus of the overlap between eigenstates of slightly different Hamiltonians)~\cite{zanardi1}, since the ground state of a quantum system changes dramatically at a critical parameter \cite{sachdev}. 

Up to now, the temporal change of fidelity -- as the overlap of the same initial states evolved by different Hamiltonians~\cite{Gorin} -- has been measured experimentally in wave billiards~\cite{fidexp2}, but also in systems of cold atoms subject to optical potentials~\cite{fidexp1,fidSAW}. Similar techniques may be applied to measure the evolving overlap of two eigenstates where time is substituted by the change of some tunable control parameter. Often a quantum phase transition may be viewed, for finite-size realizations of a system, as an avoided crossing (AC) in parameter space which closes in the thermodynamic limit \cite{sachdev}. A scenario of \emph{many} ACs with a broad distribution of widths~\cite{Haake, Kus, Wang}, as a manifestation of a strong coupling of many energy levels, is naturally found in quantum chaotic systems \cite{Haake}. The dynamical evolution of these systems is determined by the number and distribution of ACs present in the spectrum.
The question then arises whether the applicability of fidelity can be lifted from pure ground-state analysis \cite{zanardi2} to detect and characterize ACs in the entire spectrum of a complex quantum system. In this paper we propose to use the fidelity as a new and experimentally accessible tool to detect and characterize ACs in quantum spectra \cite{plotz}. This is corroborated by analytical and numerical results for exemplary quantum systems.

\section{The Fidelity measure}
\label{fidy}

Given some parameter depending Hamiltonian $H(\lambda) = H_1 + \lambda H_2$, the fidelity~\cite{Gorin} between the $n$-th eigenstates, denoted by $|n\rangle$, of two slightly different Hamiltonians $H(\lambda)$ and $H(\lambda + \delta\lambda)$ is defined as
$ f_n(\lambda, \delta\lambda) \equiv \lvert\langle n(\lambda) | n(\lambda+\delta\lambda)\rangle\rvert$. In complex quantum systems with many degrees of freedom, many of the levels of the system are coupled to each other leading to ACs in the spectrum of the Hamiltonian when the parameter $\lambda$ is changed~\cite{Haake}. To simplify the discussion, we assume a finite size Hilbert space $\mathcal H$, where all energy levels are never exactly degenerate. To detect and characterize an AC for a given quantum level $n$ we study the fidelity change~\cite{zanardi1}
	\begin{equation}\label{eq:fidchange}
		S_n(\lambda, \delta\lambda) \equiv \frac{1 - f_n(\lambda, \delta\lambda) }{(\delta\lambda)^2}  
	\end{equation}
which measures the change of the state $|n\rangle$. For $\delta\lambda\ll1$, it is independent of $\delta\lambda$, i.e. $S_n(\lambda, \delta\lambda)  \approx S_n(\lambda)$,  and vanishingly small everywhere except in the vicinity of an AC. The independence of $\delta\lambda$ arises from the fact that the first non-vanishing contribution to $f_n$ in the expansion of the changed state $|n(\lambda +\delta\lambda )\rangle$ is of second order in $\delta\lambda$ \cite{plotz,you}.
The fidelity measure (\ref{eq:fidchange}) also has the advantage of being applicable locally in the spectrum, where one follows a certain state $|n(\lambda)\rangle$ and its neighbors over a range of parameter values $\lambda$ to study the ACs they encounter. 
In addition, it is well-suited for numerical computations, since $\lambda$ is the only relevant parameter as long as $\delta\lambda$ is sufficiently small.
The different limit of large $\delta\lambda$ and hence the coupling over a broad energy band was the focus of a recent work using another generalized fidelity~\cite{hiller}. In contrast, our interest here is the detection and characterization of ACs as local couplings in energy space. 

\subsection{Two-state model}
Let us first discuss an isolated AC which can locally be described in nearly-degener\-ate perturbation theory as an effective two-level system. It is then represented by a Hamiltonian $H(\lambda) =  \lambda\sigma_z + g \sigma_x$, with a real coupling $g$ between the levels ($\sigma_x$ and $\sigma_z$ denote Pauli matrices), showing an AC at $\lambda = 0$ of width $c=2g$. The eigenstates are easily found~\cite{Haake,Landau} and from them we calculate the fidelity for the two-level system:
\begin{multline}\nonumber
f_{\pm} (\lambda,\delta\lambda) = \\
\frac{g^2+\lambda(\bar{\lambda}-\lambda)+\lambda^2+\bar{\lambda}\sqrt{g^2+\lambda^2}+\lambda\sqrt{g^2+\bar{\lambda}^2} + \sqrt{[g^2+\lambda^2][g^2+\bar{\lambda}^2]}}{2 \sqrt{\left[g^2+\lambda  \left(\lambda\pm\sqrt{g^2+\lambda ^2}\right)\right]
\left[g^2+\bar{\lambda} \left(\bar{\lambda} \pm \sqrt{g^2+\bar{\lambda}^2} \right)\right]}},
\end{multline}
where we used the shorthand notation $\bar{\lambda} \equiv \lambda + \delta\lambda$. To obtain the fidelity change in the limit $\delta\lambda\ll1$, we need to expand the expression for the fidelity in a power series for $\delta\lambda$ and keep only the leading term proportional to $(\delta\lambda)^2$. The final expression is the same for both eigenstates (indexed by $\pm$) and has the simple form:
\begin{equation}\label{eq:S2x2}
	S_{\pm}(\lambda)= \frac{1}{8}\left(\frac{g}{g^2+\lambda^2}\right)^2.
\end{equation}
This is the square of a Lorentzian and differs significantly from zero only near the AC at $\lambda = 0$. This formula already allows us a good understanding of isolated ACs, as, for example, the peak width is easily computed as $\sigma^{\text{FWHM}} = 2g\sqrt{\sqrt{2}-1}$. On the other hand an AC can be characterized by the ratio between the local energy level curvature and the distance between the two repelling energy levels. We call the absolute value of this ratio \emph{renormalized curvature} $C_n(\lambda)$ and find 
\begin{equation}\label{eq:curvature}
C_{\pm}(\lambda) \equiv \bigg|\frac{1}{\Delta(\lambda)}\frac{\partial^2 E_{\pm}(\lambda)}{\partial\lambda^2}\bigg| = 4S_{\pm}(\lambda)
\end{equation}
for the two-level system. For higher-dimensional systems we expand the wave function $|n(\lambda+\delta\lambda)\rangle$ in second order in $\delta\lambda$ and find 
\begin{equation*}
	S_n(\lambda) = \frac{1}{2} \sum_{m\neq n} \frac{\lvert\langle m(\lambda) | H_2 | n(\lambda)\rangle\rvert^2}{[E_n-E_m]^2} \approx \frac{\lvert\langle n'(\lambda) | H_2 | n(\lambda)\rangle\rvert^2}{2\;[E_n-E_{n'}]^2},
\end{equation*}
where we reduced the sum near an isolated AC to the nearest neighboring level $n'$. Similarly, one obtains for the renormalized curvature~\cite{Pechukas}
\begin{multline}\label{eq:curv2}
	C_n(\lambda) = \bigg|\frac{2}{\Delta(\lambda)} \sum_{m\neq n} \frac{\lvert\langle m(\lambda) | H_2 | n(\lambda)\rangle\rvert^2}{E_n-E_m}\bigg| \\ \approx 2\frac{\lvert\langle n'(\lambda) | H_2 | n(\lambda)\rangle\rvert^2}{[E_n-E_{n'}]^2} = 4S_n(\lambda).
\end{multline}
The relation $C_n\approx 4S_n$ thus holds as long as the effect of other levels can be neglected close to a single AC. 

\subsection{Beyond the two-level approximation} 

ACs in higher dimensional systems are not totally isolated, but other levels can contribute to the evolution of a quantum state as the parameter $\lambda$ is varied. Consider two energy levels approaching each other as $\lambda\rightarrow0$, and a third level being well separated by a distance $\epsilon$ in energy and weakly coupled to the first two levels. A Hamiltonian model for such a situation reads
\begin{equation}
	H(\lambda) = \left(\begin{array}{ccc} -\lambda & g_{} & g_{13} \\ g_{} & \lambda & g_{23} \\ g_{13} & g_{23} & \epsilon
		\end{array}\right),\quad g_{ij}, \epsilon \in \mathbb{R},
\end{equation}
where we limited ourselves to real couplings. Since the first two levels become nearly degenerate and are well-separated from the third one, we can write this in degenerate perturbation theory \cite{Shirley} close to the crossing as
\begin{equation}\label{eq:PT}
	H_{\mathrm{PT}}(\lambda) = 
	\left(\begin{array}{cc} -\lambda +\tfrac{g_{13}^2}{\epsilon} & g_{}+\tfrac{g_{13}^{}g_{23}^{}}{\epsilon} \\
 g_{}+\tfrac{g_{13}^{}g_{23}^{}}{\epsilon} & \lambda +\tfrac{g_{23}^2}{\epsilon} 	\end{array}\right) +\mathcal O (\epsilon^{-2}).
\end{equation}
This reduces the three-level system to an effective two-level system taking the effect of the distant level perturbatively into account. The same procedure can be applied, in principal, to higher dimensional systems. The minimal distance $c$ between the two levels of eq.~(\ref{eq:PT}) is thus changed by the influence of the distant third level in first order to
\begin{equation}
 	c_{\mathrm{PT}} = 2|g_{}^{}|\sqrt{\left(1+\tfrac{g_{13}^{}g_{23}^{}}{2g_{}^{}\epsilon} \right)^2+\left( \tfrac{g_{23}^2 - g_{13}^2}{2g_{}^{}\epsilon}\right)^2} 
	\approx 2|g_{}^{}|\left(1+\frac{g_{13}^{}g_{23}^{}}{2g_{}^{}\epsilon} \right),
\end{equation}
where we kept only the leading order behaviour. The minimal distance in an isolated AC is accordingly only slightly changed, provided that the coupling to the third level is not much larger than between the two encountering levels and that the third level is well-separated from  them. We need to compute the eigenstates $\ket{E_{\pm}(\lambda+\delta\lambda,\epsilon)}$ of eq.~\eqref{eq:PT} and then take their overlap for slightly different parameter values to obtain the fidelity, i.e., $f_{\pm}(\lambda,\delta\lambda,\epsilon)= |\bra{E_{\pm}(\lambda,\epsilon)} E_{\pm}(\lambda+\delta\lambda,\epsilon)\rangle|$. The fidelity change can be computed by taking the second derivative of the fidelity at $\delta\lambda=0$. The full expression is very long and difficult to grasp. Expanding it in inverse powers of $\epsilon$ and including just the first order correction to the simple two-level system, the fidelity change under the influence of a third not too close level is then given by
\begin{equation}\nonumber
S^{\mathrm{PT}}_{\pm}(\lambda,\epsilon) = \frac{1}{8}\frac{g_{}^2}{\left(g_{}^2+\lambda ^2\right)^2}
	\left[	1-\frac{2}{\epsilon} \frac{\big(g_{}^{}g_{13}^{}+\lambda  g_{23}^{}\big) 
	\big(g_{}^{}g_{23}^{}-\lambda  g_{13}^{}\big)}{g_{}^{}(g_{}^2+\lambda ^2)} +\mathcal O\left(\epsilon^{-2}\right)	\right].
\end{equation}
The correction due to the third level is also $\lambda$-dependent and changes the peak height at $\lambda=0$. Let us also include the second order correction to the fidelity change at $\lambda=0$ here
\begin{equation}\nonumber
	S^{\mathrm{PT}}_{\pm}(\lambda=0,\epsilon) = \frac{1}{8 g_{}^2}\bigg[1-\frac{2}{\epsilon}\frac{g_{13}^{}g_{23}^{}}{g_{}^{}} 
	-\frac{1}{2\epsilon^2}\frac{g_{13}^4-8g_{13}^2g_{23}^2+g_{23}^4}{g_{}^2} 
	+\mathcal O(\epsilon^{-3})\bigg].
\end{equation}
If all off-diagonal matrix elements are of similar magnitude, the effect of the third level is characterised by its inverse distance to the AC. This underlines our claim that the effect of a third level on an AC is not too strong, provided that the level is not very close. But the latter does not take place when three levels undergo a joint AC, i.e., if there were no off-diagonal matrix elements coupling the levels they would all cross in one point. Such a situation cannot be reduced to an effective two-level system. We will in the following also study numerically the behaviour of the fidelity change in exactly this case, where the third level cannot be considered a simple perturbation to the two-level system, i.e., when the approximation of an \emph{isolated} AC breaks down. 

\begin{figure}[ht!]
	\begin{center}
	\includegraphics[width = 0.95\linewidth]{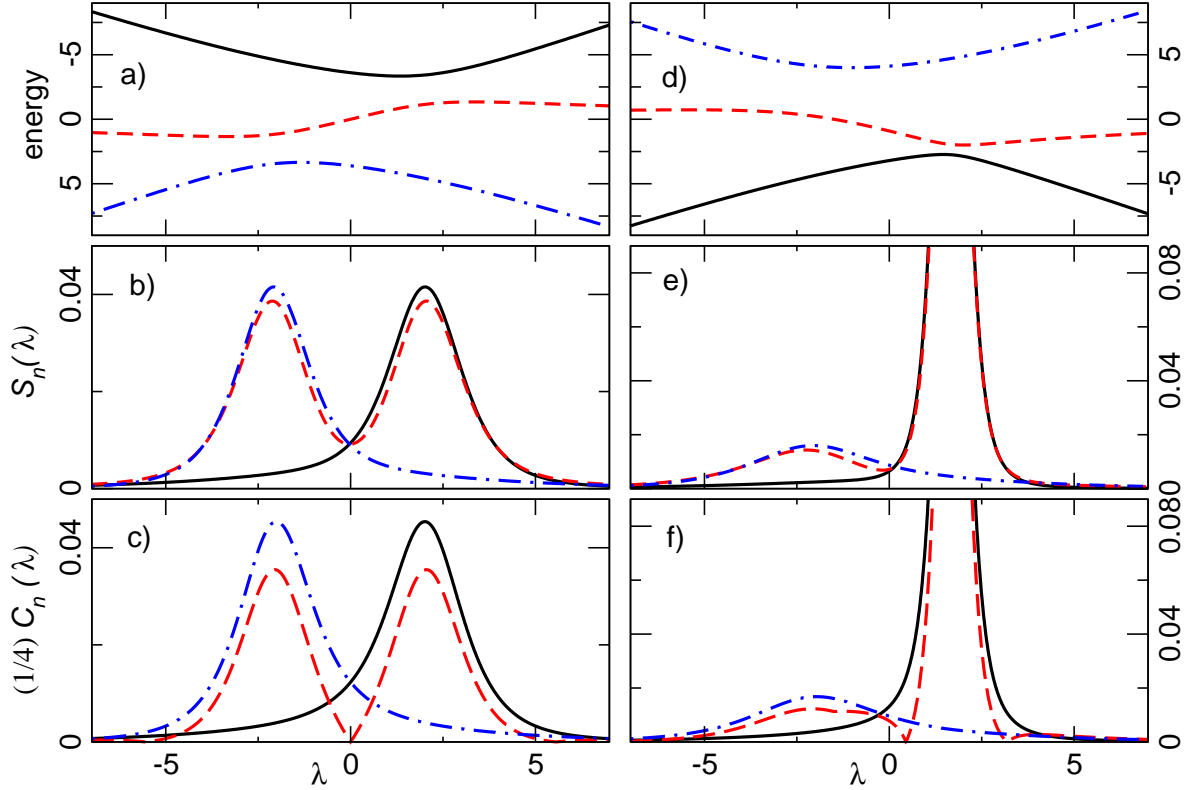}
	\caption{\label{fig:tripleAC} (a) Energy spectrum of Eq.~(\ref{eq:tripleAC}) for $a=0, b =2, c=3$. All levels are coupled and the spectrum shows two close ACs; (b) fidelity change $S_n(\lambda)$ and (c) renormalized curvature $C_n(\lambda)$ for the energy levels of (a). (d) Energy spectrum for $a=1, b =2, c=3$. All three levels are now directly coupled and the spectrum shows two close ACs. (e) $S_n(\lambda)$ and (f) $C_n(\lambda)$ for the energy levels of (d).}
	\end{center}
\end{figure}

Three crossing levels can be generated, e.g., by the following real symmetric Hamiltonian 
\begin{equation}\label{eq:tripleAC}
	H(\lambda) = \left(\begin{array}{ccc} -\lambda & a & b \\ a & 0 & c \\ b & c & \lambda\end{array}\right),
\end{equation}
which generalizes the above 2$\times$2-model. Fig.~\ref{fig:tripleAC} shows that the fidelity change, defined in Eq.~(\ref{eq:fidchange}), is able to detect and to distinguish two nearby ACs in this system. Furthermore it reflects specific features of an AC in the shape of its peak, i.e., depending on the coupling $g$, $S_n(\lambda)$ shows a narrow peak of height $S(\lambda=0)=1/(8g^2)$.

We see already in this simple example that the renormalized curvature captures the form of the fidelity change $S_n(\lambda)$ close to an AC, with deviations arising from the admixture of a further level, which first and foremost affects the local curvature, i.e., the numerator in Eq.~(\ref{eq:curvature}). But it also demonstrates that the fidelity change $S(\lambda)$ itself is still effective in detecting and characterizing the ACs. 

\section{Application to complex systems}

\subsection{Quantum chaos model}

\begin{figure}[t!]
	\begin{center}
	\includegraphics[angle=-90,width = 0.975\linewidth]{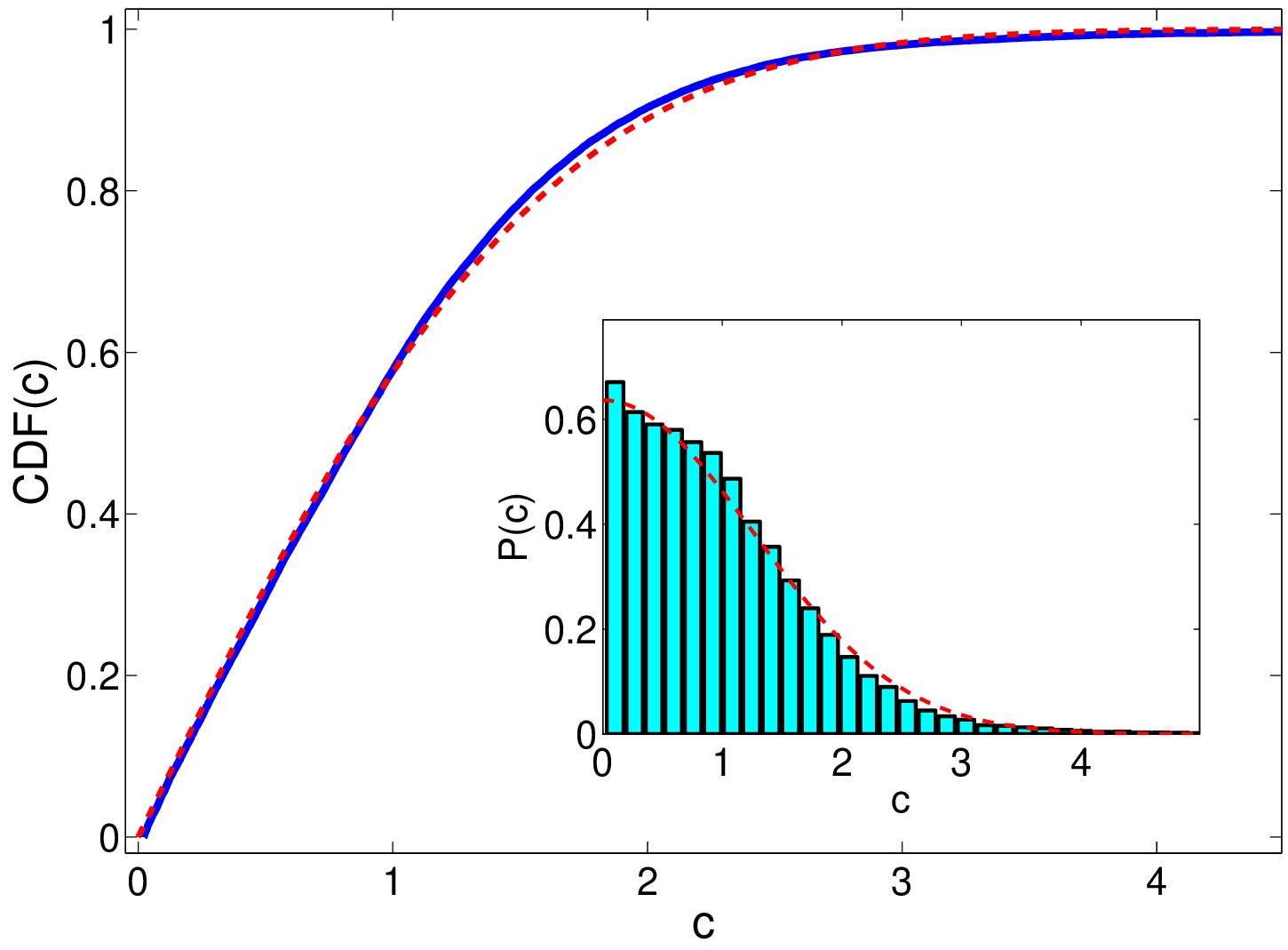}
	\caption{\label{fig:example} Cumulative distribution of ACs determined from the fidelity change maxima for the RMT model of Eq.~(\ref{eq:RMT}) with $\dim H_{1,2} = 1024$ and $\lambda \in [0,\pi[$ showing ca. 30,000 ACs: the numerical distribution (solid line) in excellent agreement with the RMT prediction CDF$(c) = \text{erf}(c/\sqrt{\pi})$ (dashed line). \emph{Inset:} Distribution of widths of the ACs $P(c)$ (histogram) and the RMT prediction (dashed line). 
	}
	\end{center}
\end{figure}

A highly dense spectrum with many and possibly overlapping ACs is encountered in quantum chaotic systems as described by Random Matrix Theory (RMT)~\cite{Haake}. A prime example having such a dense complex spectrum is the combination of two random matrices drawn from the Gaussian orthogonal ensemble~(GOE)~\cite{Haake}
\begin{equation}\label{eq:RMT}
	H(\lambda) = \cos (\lambda) H_1 + \sin(\lambda) H_2, \quad H_1,H_2 \in \text{GOE}.
\end{equation}
The distribution of minimal distances $c$  at the ACs (normalized to unit mean) is then given by a Gaussian distribution $P(c) = (2/\pi) \exp{[-c^2/\pi]}$~\cite{Kus}. Using our fidelity measure, we can directly detect the ACs in this system (by a numerical search for maxima of the $S$-function) and estimate also their widths. In the vicinity of a local maximum, the $S$-function has a Lorentzian shape as in Eq.~(\ref{eq:S2x2}) even in very dense quantum chaotic spectra. Under this assumption, we can thus extract the width of the AC as  $c = 2g=1/\sqrt{2 S_{\text{max}}}$, c.f. Eq.~(\ref{eq:S2x2}), from the local maximum $S_{\text{max}}$. Averaging over many ACs, the fidelity allows the verification of the RMT prediction with high accuracy. This is demonstrated in Fig.~\ref{fig:example} for large random matrices. 

\subsection{Bose-Hubbard system}

To further exemplify the value of our fidelity measure, we apply it to a one-dimensional Bose-Hubbard Hamiltonian with additional Stark force \cite{KB2,TMW2,ploetz}. This example of a many-body  Wannier--Stark system can be realized with ultracold atoms in optical lattices and the relevant parameters may be changed using well-known experimental techniques~\cite{BlochZwergerReview}. This model describes $N$ particles on $L$ lattice sites, with hopping between adjacent sites and a local on-site interaction. As exemplified in~\cite{KB2,TMW2}, a gauge transformation into the force accelerated frame of reference turns a constant Stark force into a time-dependent phase $\exp(\pm \ii Ft)$ with periodicity $T_B = 2\pi/F$ (the Bloch period). The corresponding Hamiltonian reads
	\begin{equation}\label{eq:BHS}	
		H(t) = - \frac{J}{2}\sum_{l=1}^{L} (\text{e}^{\iii Ft} a_{l+1}^{\dagger}a_{l} + \text{h.c.}) + \frac{U}{2} \sum_{l=1}^{L} n_l(n_l-1), 
	\end{equation}
where $a_l^{\dagger}$ ($a_l$) creates (annihilates) a boson at site $l$ and $n_l = a_l^{\dagger}a_l$ is the number of bosons at site $l$. The parameter $J$ is the hopping matrix element, $U$ the interaction energy for two atoms occupying the same site, and $F$ the Stark force.
Periodic boundary conditions are imposed for $H(t)$, such that the Hamiltonian and the one-period Floquet operator 
	$\hat U_F(T_B) = \mathcal T \text{exp} \left(-i\int_0^{T_B} H(t) \text{d} t \right)$ 
(where $\mathcal T$ denotes time-ordering) decompose into a sum of operators for specific quasimomenta $\kappa$~\cite{KB2}.
In the following, we use $F$ as a control parameter. For $J \approx U \ll F$ the quasienergy spectrum (eigenphases of $\hat U_F(T_B)$) is dominated by the force $F$ and the system is regular. Decreasing the force to $J \approx U \gtrsim F$ the quasienergy spectrum reorders and the coupling between the levels becomes more important. For fillings of order unity, e.g. $N/L \approx 1$, the system is quantum chaotic in this regime and the spectrum obeys Wigner-Dyson statistics~~\cite{KB2, TMW2}. As $F$ is varied one observes an increasing number of ACs as the spectrum is changing and additionally many broad ACs once the quantum chaotic region is reached. 

\begin{figure}[t!]
	\begin{center}
	\includegraphics[angle=-90, width = 0.94\linewidth]{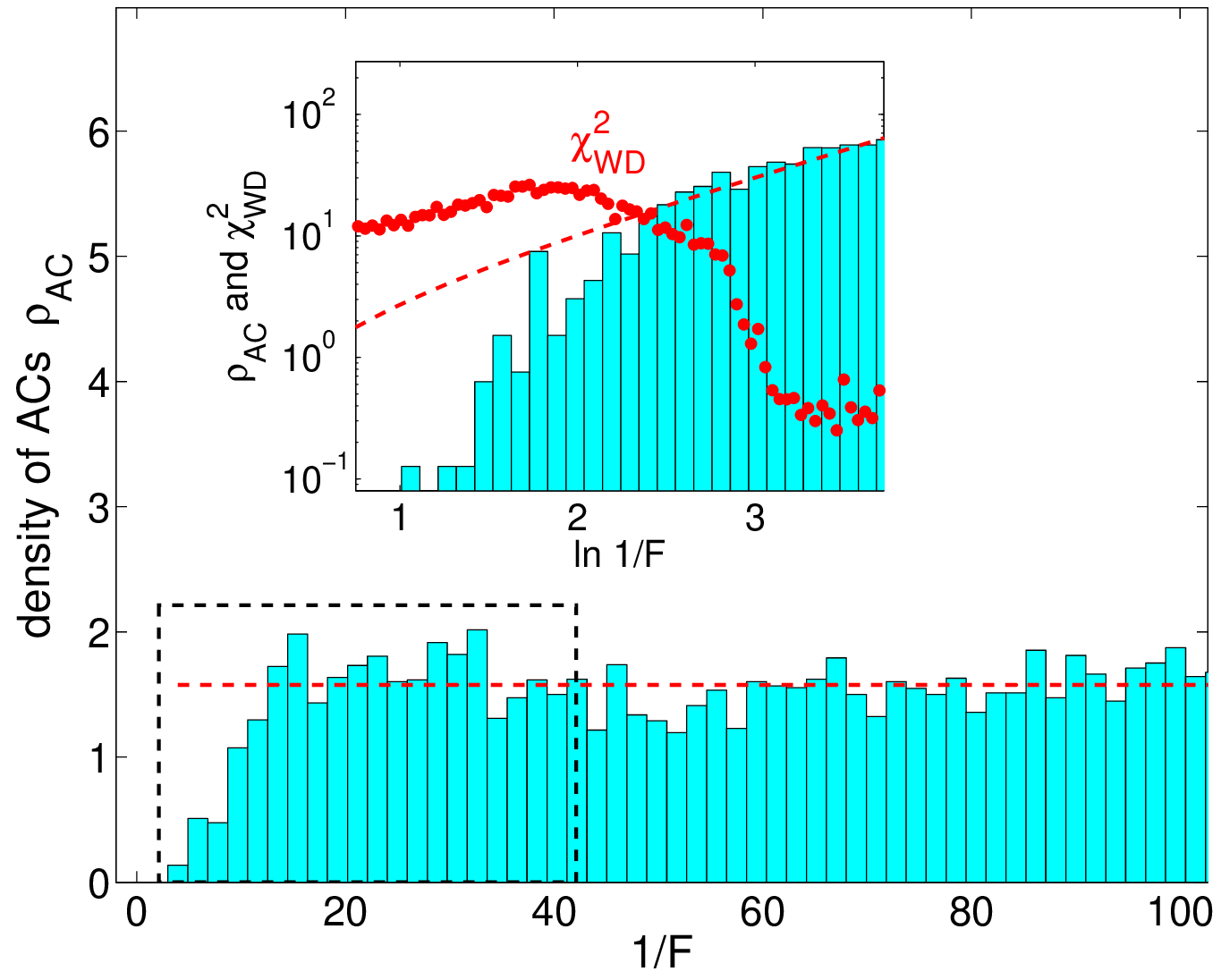}
	\caption{\label{fig:tBH} Density of ACs $\rho_\text{AC}$ in the quasienergy spectrum of the Floquet operator of our Bose-Hubbard model for varying $F$ and fixed $J=0.038$, $U=0.032$, $N=L=6$. The number of ACs as detected by the fidelity change increases with $1/F$ and saturates around $1/F\approx 20$ to an average value which is shown by the dashed line. {\em Inset:} Magnification of the region marked by the box on logarithmic scale with a comparison to a $\chi^2$ test (with small values for good Wigner-Dyson statistics \cite{TMW2}).}
	\end{center}
\end{figure}

To illustrate the crossover between regions with few and many ACs, we study the density of ACs as detected by the  fidelity change $S_n$, when changing the system parameter $\lambda$. In a histogram, the density $\rho_{\text{AC}}(\lambda)$ is defined via
	$\rho_{\text{AC}}(\lambda)\cdot d\lambda \equiv N_{\text{AC}}(\lambda)/\text{dim}\mathcal H,$
comparing the number of ACs $N_{\text{AC}}(\lambda)$ in the interval $[\lambda, \lambda+d\lambda]$ to the total number of energy levels $\text{dim}\mathcal H$. This is shown in the main part of Fig.~\ref{fig:tBH} where we observe no ACs at large $F$, i.e., small values of $1/F$, and an increasing number of ACs for larger values of $1/F$ that saturates around $1/F\approx 20$. 

The mentioned transition between regular and chaotic spectral properties for $J \approx U \gtrsim F$ and approximately integer filling in the tilted system can be visualized by comparing the actual level spacing distribution to a Wigner-Dyson distribution using a standard statistical $\chi^2$ test~\cite{TMW2}. This is displayed in the inset of Fig.~\ref{fig:tBH} along with the density of ACs in Fig.~\ref{fig:tBH}. The fidelity change $S(1/F)$ detects ACs and shows the same qualitative behavior as the spectral statistics along the crossover from regular to chaotic dynamics: in regions of good Wigner-Dyson statistics we find a high density of ACs compared to a smaller number of ACs in the regular regime. The crossover beginning for $\log(1/F) \approx 2$, where the density of ACs rises above unity, i.e., on average each energy level undergoes more than one AC in the unit interval. The transition is complete for $\log(1/F) \approx 3$ where the $\chi^2$ test saturates around a low value. However, the density of ACs alone is not able to distinguish regular from chaotic dynamics. Instead the ACs need to have a broad distribution of widths which is reflected in the distribution $P(c)$ introduced above.

\begin{figure}[t!]
	\begin{center}
	\includegraphics[angle =-90, width = 0.99\linewidth]{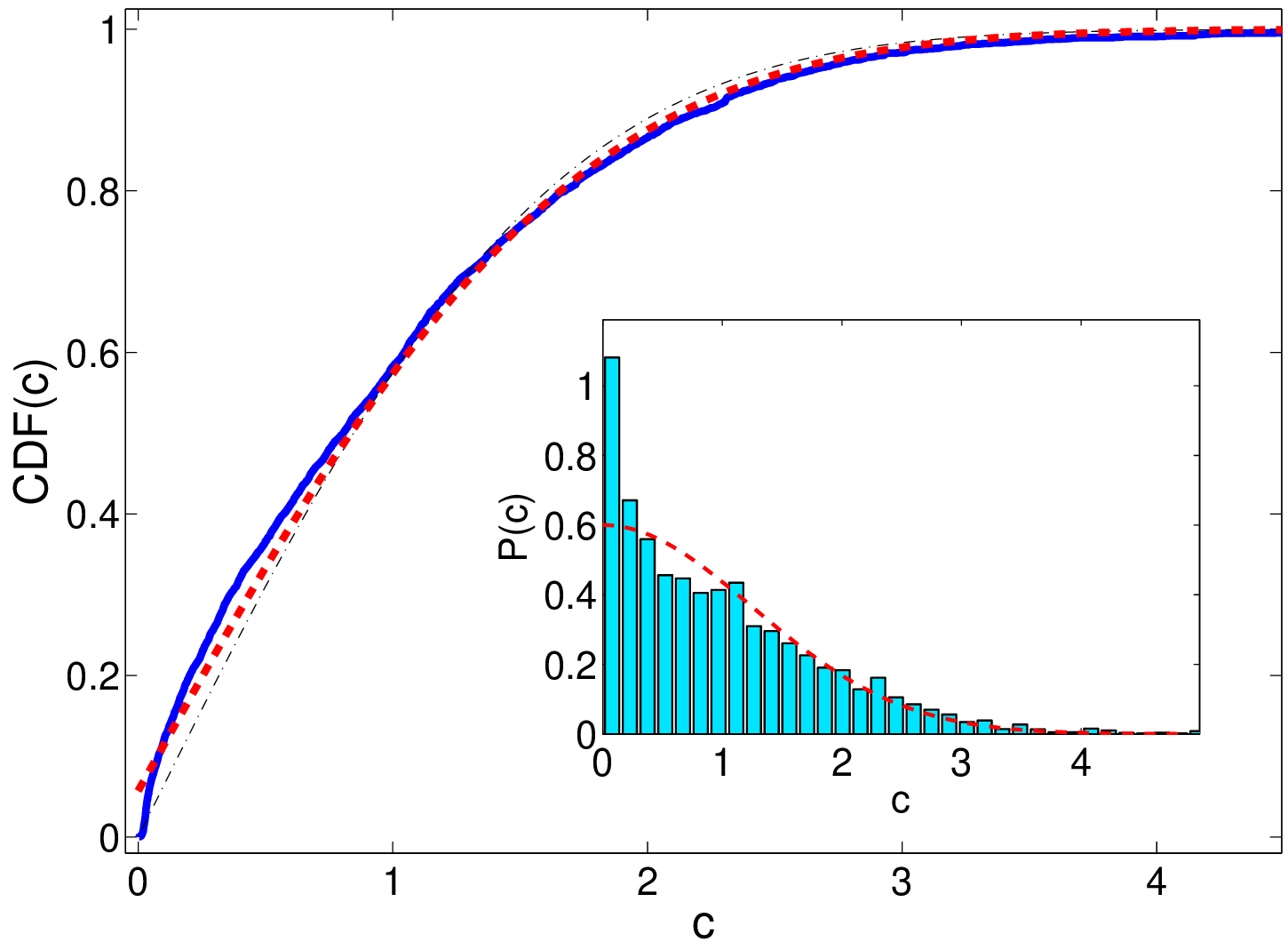}
	\caption{\label{fig:BH} Cumulative distribution of ACs determined from the fidelity change maxima for the system of Eq.~(\ref{eq:BHS}). Shown are the numerical distribution (solid line), the best fit for a mixed RMT spectrum (thick dashed line, chaotic part $\gamma \approx 0.94$), and the RMT prediction for a purely chaotic spectrum (thin dashed-dotted line). Parameters: $N=6, L=7, J = 0.038, U = 0.032, F = 1/39 \ldots 1/35$.
	\emph{Inset:} Distribution of widths of ACs for the same model (histogram) and Eq.~\eqref{eq:fit} with $\gamma \approx 0.94$ (dashed line). The enhancement close to $c=0$ arises from regular ``solitonic'' states \cite{soliton} in the spectrum.}
	\end{center}
\end{figure}

By using the fidelity change in order to detect and characterize ACs, we can resolve further remarkable details in the full spectrum.  With this method we are, e.g., able to detect a small number of regular states~\cite{soliton} traversing the chaotic sea of energy levels in the chaotic regime of the tilted Bose--Hubbard model. In this case the distribution of widths of ACs is a mixture of regular and quantum chaotic distributions:
\begin{equation}\label{eq:fit}
	P(c) = (1-\gamma) \delta(c) + \frac{2\gamma^2}{\pi \bar c}\;\text{exp}\left[-\frac{\gamma^2c^2}{\pi \bar c^2}\right],
\end{equation}
with a chaotic part of weight $0\leq\gamma\leq1$~\cite{burgdoerfer}. A finite regular component makes itself visible as a strong enhancement of $P(c)$ close to zero, c.f., the inset of Fig.~\ref{fig:BH}. 
We are able to estimate the size of this component by analyzing the cumulative distribution function $\text{CDF}(c) = 1-\gamma +\gamma\; \text{erf}\left(\frac{\gamma c}{\sqrt{\pi}} \right)$. The result is shown in the main part of Fig.~\ref{fig:BH}, where we plot the numerically obtained distribution and the best $\chi^2$-fit including a finite regular component. We obtain a chaotic part of $\gamma \approx 0.94$, corresponding to ca. 6\% of regular levels, in good agreement with counting 7 regular levels out of 132 by direct inspection of the spectrum. Except for the identification of single regular levels~\cite{soliton}, this has so far not been detected in the tilted Bose--Hubbard model by other statistical measures. The reported results are obtained for periodic boundary conditions applied to the Hamiltonian of Eq.~(\ref{eq:BHS}), but we found a qualitatively similar picture for hard-wall boundary conditions, as used in \cite{soliton}. Our results underline the value of fidelity as a measure for detecting ACs with high resolution in energy spectra.

\section{Conclusions}

We showed that quantum fidelity is perfectly suited to detect and characterize ACs in the energy spectrum. It therefore connects information about the wave function of a system with its spectrum, without direct reference to the energy levels by using only the overlap of wave functions \cite{Kohler}. This has been exemplified for simple models and for complex quantum systems showing many ACs. The fidelity, therefore, proves very useful to study many-body systems, also beyond their ground-state properties \cite{zanardi1}. 

We expect a clear advantage of the fidelity change compared to spectral statistics in the sense that it can be applied, in principle, also just \emph{locally} in the spectrum. This means that, if one is interested only in local spectral properties of a system, it is sufficient to follow a small number of levels to characterize the behavior of a system. For larger systems, computing the entire spectrum and all eigenstates is in general difficult, but the fidelity allows an analysis of parts of the spectrum providing local spectral information. To make use of this advantage, one may resort to numerical algorithms optimized to access just a subset of eigenstates, e.g., the Lanczos algorithm \cite{lanczos}. We will pursue this interesting perspective of the fidelity proposed here in a future publication \cite{carlos}.

\section{Acknowledgments}

This work was supported by the HGSFP (DFG grant GSC 129/1), FOR760 (DFG grant WI 3426/3-1), the Frontier Innovation Fund, the Global Networks Mobility Measures, and the Klaus Tschira Foundation. We thank R\'emy Dubertrand, Boris Fine, Andrea Tomadin, and especially Steve Tomsovic for many inspiring discussions.

\bibliographystyle{elsarticle-num}



\end{document}